\documentclass[a4paper,floatfix,superscriptaddress,pra,twocolumn,longbibliography]{revtex4-2}

\usepackage{lineno}
\usepackage{dsfont}
\usepackage{times,bbm,amsmath,amssymb,mathtools}
\usepackage{epsfig,color}
\usepackage{xcolor}
\usepackage{hyperref}
\hypersetup{
    colorlinks = true
}
\usepackage{cleveref}
\usepackage{microtype}
\usepackage{float,siunitx}
\usepackage[caption = false]{subfig}
\usepackage[greek,english]{babel}
\usepackage{thumbpdf,enumerate}
\usepackage{booktabs}
\usepackage{sidecap}
\usepackage[scaled=.8]{couriers}
\usepackage{multirow}
\usepackage{placeins}
\usepackage{relsize}
\usepackage{pst-grad,bm}
\usepackage{epigraph}
\usepackage{longtable}
\usepackage{ulem} 
\usepackage{acronym}
\usepackage{physics}
\usepackage{easyReview}

\begin{document}

\title{Photonic cellular automaton simulation of relativistic quantum fields: observation of Zitterbewegung}

\author{Alessia Suprano}
\email{These authors equally contributed to this work.}
\author{Danilo Zia}
\email{These authors equally contributed to this work.}
\address{Dipartimento di Fisica, Sapienza Universit\`{a} di Roma, Piazzale Aldo Moro 5, I-00185 Roma, Italy}
\author{Emanuele Polino} 
\address{Dipartimento di Fisica, Sapienza Universit\`{a} di Roma, Piazzale Aldo Moro 5, I-00185 Roma, Italy}
\address{Centre for Quantum Dynamics and Centre for Quantum Computation and Communication Technology,
Griffith University, Brisbane, Queensland 4111, Australia}
\author{Davide Poderini} 
\address{Dipartimento di Fisica, Sapienza Universit\`{a} di Roma, Piazzale Aldo Moro 5, I-00185 Roma, Italy}
\address{International Institute of Physics, Federal University of Rio Grande do Norte, 59078-970, P. O. Box 1613, Natal, Brazil}
\author{Gonzalo Carvacho}
\author{Fabio Sciarrino}
\email{fabio.sciarrino@uniroma1.it}
\address{Dipartimento di Fisica, Sapienza Universit\`{a} di Roma, Piazzale Aldo Moro 5, I-00185 Roma, Italy}
\author{Matteo Lugli}
\author{Alessandro Bisio}
\author{Paolo Perinotti}
\email{paolo.perinotti@unipv.it}
\address{QUIT Group, Dipartimento di Fisica, Università degli Studi di Pavia, and INFN sezione di Pavia, via Bassi 6, 27100 Pavia, Italy}




\vspace{10pt}

\begin{abstract}
Quantum Cellular Automaton (QCA) is a model for universal quantum computation and a natural candidate for digital quantum simulation of relativistic quantum fields. Here 
we introduce the first photonic platform for implementing QCA-
simulation of a free relativistic Dirac quantum field in 1+1 dimension, through a Dirac 
Quantum Cellular Automaton (DQCA). Encoding the field position degree of freedom in the Orbital Angular Momentum (OAM) of single photons, our state-of-the-art setup experimentally realizes 8 
steps of a DQCA, with the possibility of having complete control over the input OAM state preparation and the output measurement making 
use of two spatial light modulators. Therefore, studying the distribution in the OAM 
space at each step, we were able to reproduce the time evolution of the free Dirac field observing, the \textit{Zitterbewegung}, an oscillatory movement extremely difficult to see in real case experimental scenario that 
is a signature of the interference of particle and antiparticle states.
The accordance between the expected and measured \textit{Zitterbewegung} oscillations 
certifies the simulator performances, paving the way towards 
the application of photonic platforms to the simulation of more complex relativistic effects. 
\end{abstract}

\maketitle 

\section{Introduction}
The notion of a cellular automaton was introduced by von Neumann \cite{neumann1966theory}, with the purpose of showing how a simple local update rule for an array of cells containing bits (or larger information carriers) can produce complex behaviours on a macroscopic scale. The quantum version of cellular automata, Quantum Cellular Automaton (QCA)~\cite{schumacher2004reversible,gross2012index,arrighi2011unitarity,farrelly2020review,freedman2020classification} was first envisaged by Feynman in his famous paper~\cite{feynman2018simulating}, where he immediately proposes their use as quantum simulators. A QCA consists of a lattice of finite-dimensional quantum systems, along with an evolution occurring in discrete steps, which can be summarized in a local update rule, i.e.~involving only a finite neighbourhood in the update of a given cell (see Fig.~\ref{fig:QW_figurelocality}). 

Recently, quantum cellular automata have attracted great interest due to their potential in quantum computation \cite{Watrous1995QCA,Raussendorf2005QCA,Vollbrecht2006QCA} and because they are universal digital quantum simulators~\cite{Arrighi_2012,Chun2016Floquet,Fidkowski2019Floquet,Osborne2006Efficient,Molfetta2013Quatum, Cedzich2013Propagation, DIMOLFETTA2014157, D_Ariano_2015,bisio2021scattering,PhysRevLett.125.190402}. In this context, special attention was paid to the single particle sector of QCAs whose dynamics reduces to a Discrete Time Quantum Walk (DTQW) \cite{ambainis2001one}. DTQWs are useful for quantum computation \cite{shenvi2003quantum,kempe2003quantum,ambainis2005coins,portugal2013quantum} and simulation \cite{Zhang,cardano2016statistical,Kitagawa,Xiao_2017,Cardano_2017, Zhan_2017}, which makes them worth to realize in their own right. 
As the single-particle sector of a QCA, DTQWs can simulate free quantum relativistic field theories, and relevant effects thereof~\cite{kurzynski2008relativistic,mallick2016dirac,bisio2013dirac,PhysRevA.90.062106,BISIO2016177,PhysRevA.94.042120}.


Several platforms ranging from cold atoms \cite{karski2009quantum}, trapped ions \cite{schmitz2009quantum, Zahringer2010}, to photonics systems \cite{sansoni2012twoparticle, giordani_2018, cardano2015quantum, suprano2021dynamical, crespi2013anderson, caruso2016fast, kitagawa2012observation, owens2011twophoton, qiang2016efficient, boutari2016large, cardano2016statistical, Cardano_2017,Esposito22}, have been employed to implement DTQWs. 
In this paper, we focus on the experimental realization of the Dirac Quantum Cellular Automaton (DQCA)~\cite{DARIANO2012697,BISIO2015244,PhysRevA.90.062106}, which is a Fermionic Cellular Automaton which recovers the dynamics of a free Dirac field in the small wave-vector regime. The digital simulation of the special instance of its evolution on  localized input states was pioneered in Ref.~\cite{alderete2020quantum} on a trapped-ion quantum computer. 

In this work, we implement the single particle sector of a QCA with the same dispersion relation as that of the DQCA, corresponding to
a DTQW, using a photonic platform based on the scheme proposed in Ref.~\cite{bisio2013dirac}. We use the Orbital Angular Momentum (OAM) of light to encode the \textit{walker} system that is directly linked to the position of the Dirac particle, while the \textit{coin} is codified in the polarization degree of freedom. The structured wavefront characterizing OAM states and their high-dimensionality motivate the wide applications that these states have found both in the classical regime, regarding microscopy \cite{furhapter2005spiral,tamburini2006overcoming}, optical trapping \cite{Zhan} and communication \cite{willner2015optical,bozinovic2013terabitscale,malik2012influence,baghdady2016multi,wang2016advances}, and in quantum information processing for the development of protocols in quantum communication \cite{Wang2015,Cozzolino_rev,cozzolino2019air}, computation \cite{Lanyon2009,ralph2007efficient}, metrology \cite{dambrosio_gear2013,fickler2012quantum,cimini2023experimental} and cryptography \cite{Mirhosseini_2015,Bouchard_18}. Moreover, OAM-based platforms offer the possibility to produce a DTQW on a line  without an exponential increase of the number of optical elements with respect to the length of the walk \cite{giordani_2018,suprano2021dynamical, cardano2015quantum, cardano2016statistical, Cardano_2017}.

Here, we move beyond the present status of experimental OAM-based quantum walk platforms, implementing 8 steps of a DTQW with a controllable initial state and an arbitrary projective measurement stage at the output. 
As a certificate of our simulation, we observed the \emph{Zitterbewegung}, a quivering motion of a relativistic particle that, despite being practically impossible to observe in relativistic systems, is considered as one of their benchmark signatures. Indeed, it was sought in the pioneering---and to date one of the very few---quantum simulations of Dirac equation in a trapped ion system~\cite{gerritsma2010quantum}. To the best of our knowledge, we achieved the first digital quantum simulation of Zitterbewegung.
%
%
%
%
%
%
Our work demonstrates the capability of photonic platforms of simulating relativistic behaviour, that is difficult to observe in real case scenarios, paving the way for further experimental implementations of QCA and DQCA, also with more complex evolution dynamics thanks to the reconfigurability of the platform for what concern input and output stages.

\section{Quantum Cellular Automata and Quantum Walks}

\begin{figure}[t]
 \centering
    \includegraphics[width=1\columnwidth]{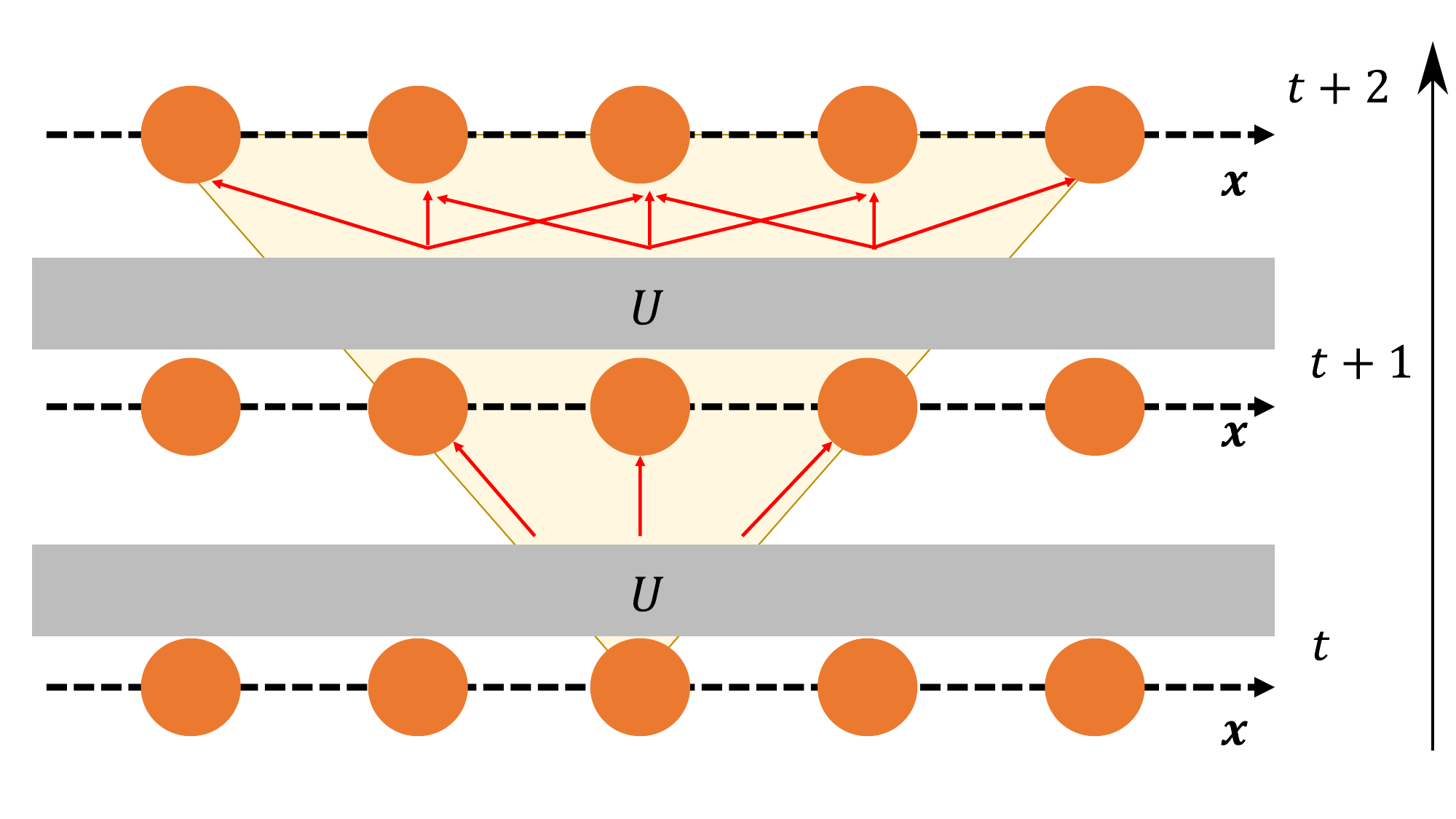}
  \caption{\textbf{One-dimensional QCA}. Representation of the one-dimensional QCA discrete time evolution described by the unitary operator \textit{U}. Here, the quantum field located at each lattice point interacts locally only with the nearest neighbors at each step of the evolution.} 
  \label{fig:QW_figurelocality}
\end{figure}

Quantum cellular automata describe the unitary evolution of a lattice of cells, each representing a quantum system.
The evolution occurs in discrete steps and it is \emph{local}, namely the state of a cell after a certain step $t+1$ 
depends only on the state of finitely many neighboring cells after the preceding step $t$ 
(see Fig.\ref{fig:QW_figurelocality}).

Let us consider the one-dimensional nearest-neighbor lattice $\mathds{Z}$ and a local Bosonic (Fermionic) mode per cell.
We associate every site $x \in \mathds{Z}$ with an algebra of field operators $\psi_{x,a}$ where the index $a\in S$ belongs to a finite set $S$ and denotes some internal degree of freedom (e.g.~polarization, spin, helicity, \ldots). The field operators fulfill either the Canonical Commutation Relations (CCR) $[ \psi_{x,a},  \psi_{y,b}^\dagger] = \delta_{x,y}\delta_{a,b}$, $[ \psi_{x,a},  \psi_{y,b}] = [ \psi_{x,a}^\dagger, \psi_{y,b}^\dagger] = 0$ or the Canonical Anticommutation Relations (CAR) $\{ \psi_{x,a}, \psi_{y,b}^\dagger \} = \delta_{x,y}\delta_{a,b} $, $\{ \psi_{x,a}, \psi_{y,b} \} = \{ \psi_{x,a}^\dagger, \psi_{y,b}^\dagger \} = 0$, $\forall x, y\in\mathds Z$ and $a,b\in S$.
A quantum cellular automaton $\mathcal{U}$ is a local and translation invariant automorphism of the representation of the CCR (CAR) algebra which represents a one-step evolution of the lattice. 
We give a Fock space representation of the CCR (CAR) algebra by introducing the $N$-excitations (particles) states
$\ket{(x_1,a_1),\ldots,(x_N,a_N)}\coloneqq  \psi^\dag_{x_1,a_1}\cdots  \psi^\dag_{x_N,a_N}\ket\Omega$,
where $\ket\Omega$ is the {\em vacuum state}, i.e the state with no excitations which obeys
$\psi_{x_i,a_i}\ket\Omega =0$ for all $i$. \footnote{We defined the vacuum
  state as the one with no \emph{localized}
  excitations. In quantum field theory one usually
  defines the vacuum state as the ground state of a
  free Hamiltonian which is a collection of
  harmonic oscillators. There,
  particles are not localized excitation but they
  have a well defined energy and momentum (as
  improper eigestates). Since we will consider the
  free evolution of a single particle, the
  difference between the two construction is
  immaterial. We chose the local excitation basis
  for 
  convenience.}
If we consider the particular case
of a free, i.e.~non-interacting, evolution, the
QCA action is linear in the field operators, namely
\begin{align}
  \label{eq:7}
  \mathcal{U}( \psi_{x,a})=\sum_{y\in\mathds Z}\sum_{b\in
  S}U^*_{y,b;x,a} \psi_{y,b},
\end{align}
where the coefficients $U_{y,b;x,a}$ turn out to be matrix elements of a unitary operator on the subspace spanned by single-particle states.

Thus, the dynamics is completely determined by the {\em quantum walk}
$U$ on the single-particle Hilbert space
$\mathds{C}^S \otimes l_2
(\mathds{Z})$
\begin{align}
  \begin{split}\label{eq:1}
 &\ket{\psi(t+1)}=U\ket{\psi(t)},\\
 &U\ket{a}\ket x=\sum_{y\in\mathds Z}\sum_{b\in S}U_{y,b;x,a} \ket{b}\ket y, \\
 &\ket a\ket x = \ket{(x,a)}.
\end{split}
\end{align}

We now consider a two-dimensional internal degree of freedom corresponding to the polarization of an e.m.~field mode. The Hilbert space $\mathds C^2$ is thus spanned by the polarization eigenstates, $\{\ket{V}, \ket{H}\}$, and the circularly polarized states are denoted as $\ket L$,$\ket R$ with 
\begin{align*}
\ket L = \frac{1}{\sqrt{2}}(\ket{V}+i\ket{H}),\quad\ket R = \frac1{\sqrt{2}}(\ket{V}-i\ket{H}).
\end{align*}

Because the evolution is translationally invariant it is convenient to represent the unitary operator $U$ in Eq.~\eqref{eq:1} through the momentum representation:
\begin{align}
\label{eq:5}
  U = \int_{-\pi}^{\pi} \!\!\!\! \text{d}k \ U(k) \otimes \ket{k}\bra{k} , \;\;
  U(k)   \ket{\pm}_k = e^{\mp i \omega(k)}\ket{\pm}_k
  \end{align}
  where we introduced the plane waves $ \ket{k} :=
  \sum_x \frac{e^{ikx}}{\sqrt{2\pi}} \ket{x}, $
  and $U(k) \in SU(2)$ is a unitary matrix with eigenvectors $\ket{+}_k$ and $\ket{-}_k$.\footnote{In general, $U(k)$ can have determinant which depends on $k$.
  However, one can prove that every DTQW can be decomposed in terms of left (right) shifts on the lattice and DTQWs such that $U(k) \in SU(2)$ for all $k$ \cite{Gross:2012uw}.}
  For example, the DTQW corresponding to the one particle sector of the Dirac Cellular Automaton~\cite{BISIO2015244,mallick2016dirac,bisio2013dirac} reads as follows:
  \begin{align}
    \label{eq:2}
      U(k) = \begin{pmatrix}
        n e^{-ik} & -im \\
        -im & n e^{ik}
      \end{pmatrix}, \;\;
      \omega(k) 
     = \arccos (n \cos(k))
  \end{align}
  for some real numbers $n,m$ such that $n^2+m^2 =1
  $.

For a given quantum walk $U$, it is useful to
introduce an \emph{effective Hamiltonian} $H$ which
obeys $U = \exp(-i H)$.
  The Hamiltonian $H$ generates a continuous
  time evolution which interpolates the evolution
  of the quantum walk. We refer to the support
  $\mathcal{H}_+$ (resp. $\mathcal{H}_-$) of
  the projector
  $P_{\pm} := \int \! \text{d}k \,
  \ketbra{\pm_k}{\pm_k}
  \otimes \ketbra{k}{k}$
  as the subspace of \emph{positive} (resp.
  \emph{negative}) \emph{energy} states.
 
  In particular, for the DTQW of Eq.~\eqref{eq:2} we have
  \begin{align}
    \label{eq:6}
    \begin{aligned}
     {H} &=   \int_{-\pi}^{\pi} \!\!\!\! \text{d}k \ H(k) \otimes \ket{k}\bra{k} , \\
  H(k) &= \frac{\omega(k)}{\sin \omega(k)}\begin{pmatrix}
        n \sin(k) & m \\
        m & - n \sin(k)
      \end{pmatrix},      
    \end{aligned}
  \end{align}
  and one can easily verify that for small $k$ and
  $m$ the
  one dimensional Dirac equation
  \begin{align}
    \label{eq:3}
    i \partial_t \psi(k,t) = (k\sigma_z + m \sigma_x) \psi(k,t)
  \end{align}
  is recovered.
  The above considerations show that the DTQW
  in Eq.~\eqref{eq:2} provides a quantum
  simulation of the one dimensional Dirac free
  field and can be used to observe relativistic
  quantum effects pertaining regimes that are
  difficult to access experimentally.

\begin{figure*}[t!]
    \centering
    \includegraphics[width=\textwidth]{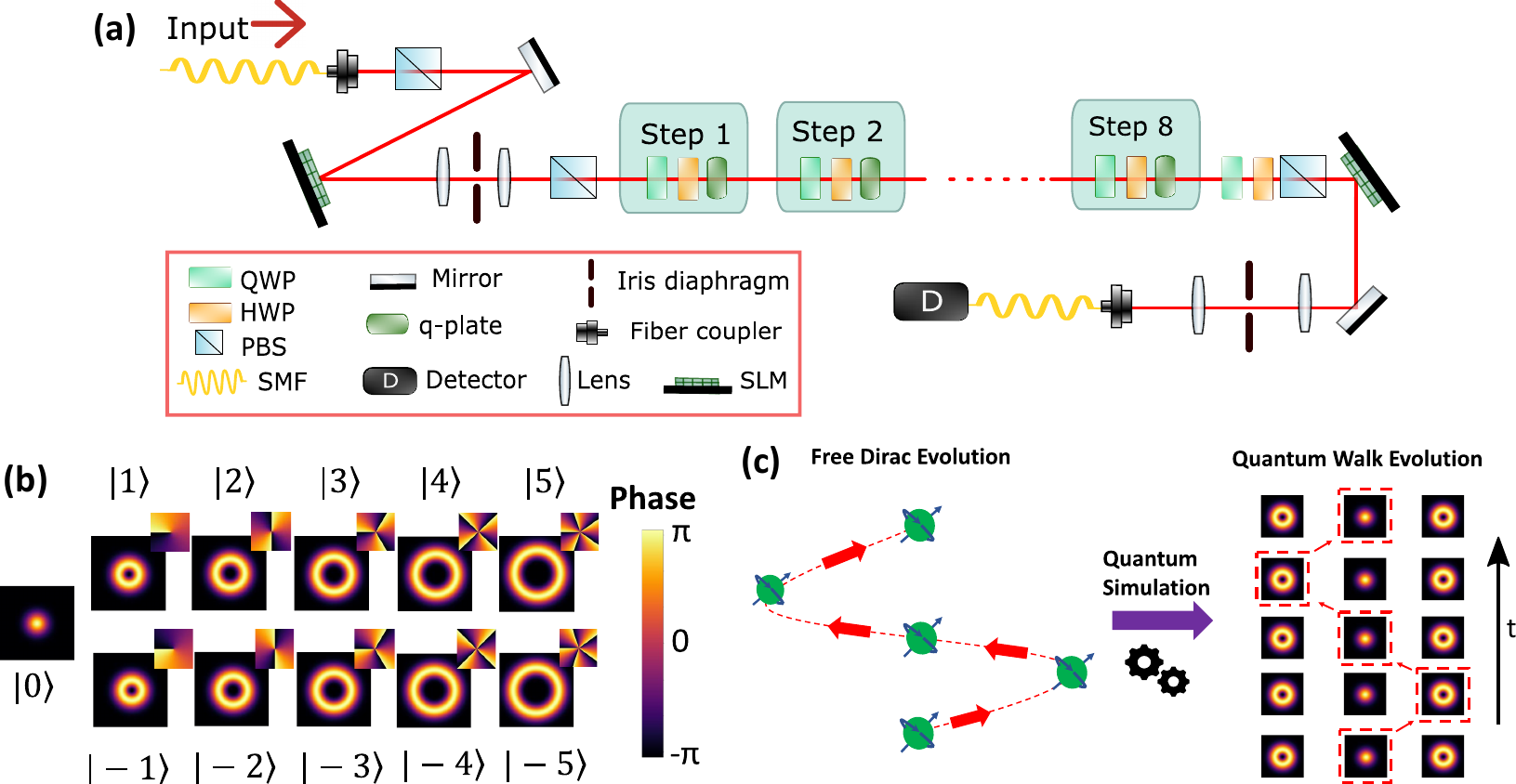}
    \caption{\textbf{Experimental setup:} (a) The Quantum Cellular Automaton evolution is implemented through an eight-step discrete-time quantum walk in the OAM of light. First of all, single photon states are generated through spontaneous parametric down-conversion in a Periodically Poled Potassium Titanyl Phosphate (PPKTP) non-linear crystal. 
    After projecting the polarization of single-photons on the horizontal one through a Polarizing Beam Splitter (PBS), the desired input state is produced via a spatial light modulator (SLM) and, after a spatial filtering performed with an iris diaphragm, is sent to the DTQW. Each step of the latter consist in a coin operator, implemented by a quarter-waveplate (QWP) and a half-waveplate (HWP), and a shift operator performed using a q-plate. Then, the polarization is traced out using a series of QWP, HWP and PBS. The output state probability distribution is measured with a projective measurement executed via a further SLM followed by a single-mode fiber (SMF), the resulting coupled signal is detected by an avalanche photodiode detector. (b) Mapping between the OAM space and the position space. In particular, each position of the Dirac particle is identified with a different OAM eingenstate. For the latter, we report both the intensity and the phase of the wave function as expressed in the Laguerre-Gaussian modes basis. (c) The time evolution of a free Dirac particle is simulated through the DTQW platform using the orbital angular momentum of single photons. Here, a modification in the particle position is identified with a variation of the OAM value. }
    \label{fig:QWP}
\end{figure*}

\subsection*{\emph{Zitterbewegung} in quantum walks}

One of the main predictions of the Dirac equation
is the existence of antiparticles. As first noticed
by Schr{\"o}dinger
\cite{schrodinger1930kraftefreie},
interference of a Dirac particle with its
antiparticle is responsible for the
so-called \emph{Zitterbewegung} effect, namely the
oscillation of the expected value of the position
operator ${X}$ \cite{thaller2013dirac,kurzynski2008relativistic,david2010general, mallick2016dirac, bisio2013dirac,lock1979zitterbewegung, ahrens2015simulation,juneghani2021study,lamata2007dirac, tran2014optical,pedernales2013quantum,zawadzki2011zitterbewegung,schliemann2005zitterbewegung,vaishnav2008observing,zhang2008observing,liang2011zitterbewegung, deng2015optically,zhang2012relativistic}.
Direct observation of this phenomenon in particle physics would be prohibitive since it requires preparing a coherent superposition of particle and antiparticle and the oscillation amplitude is of the order of the Compton wavelength ($10^{-12}$ m for an electron).

Since this phenomenon ultimately depends only on the presence of positive and negative energy states, it can be observed also in DTQWs
\cite{bisio2013dirac}. For the case of a quantum
walk on a one-dimensional lattice (see Equation~\eqref{eq:5}) the position
operator is $ {X} := \sum_{x \in \mathds{Z}} x
\, I \otimes \ketbra{x}{x}$ and its time evolution
$ {X}(t) = U^{-t}  {X} U^{t}$ can be
computed by integrating the differential equation
$\frac{\text{d}^2}{\text{d}t^2} {X}(t)  = - [H,[H, {X}]]$
where $H$ is the effective Hamiltonian.
We obtain 
\begin{align}\label{eq:X_t}
  \begin{aligned}
    &	  X(t) =   X(0) +   V t +
        \frac{1}{2i   H}
        \left(e^{2i   H t} - I\right)   F, \\
  &  {V} :=   \int_{-\pi}^{\pi} \!\!\!\!
  \text{d}k \ \frac{\omega'(k)}{\omega(k)}H(k)
  \otimes \ket{k}\bra{k} , \quad
  F := [ {H}, {X}] -  {V}.
  \end{aligned}
\end{align}
where $ {V} $ is the velocity operator and 
$  F$ 
is responsible for the oscillating motion. Since
$  {F}P_{\pm} = P_{\mp} {F} $,
we have that the Zitterbewegung
occurs only for states which are a
superposition of positive energy (particle) and negative energy
(antiparticle) states.
Indeed by taking the expectation value of $ {X}(t)$
with respect to a state 
$    \ket{\psi} = \ket{\psi}_+ +
  \ket{\psi}_-$, where  $\ket{\psi}_{\pm} \in
  \mathcal{H}_{\pm}$,
we have
\begin{align}
  \label{eq:8}
  \begin{aligned}
  \langle X(t)\rangle &= x_+(t) + x_-(t) + x_0 + z (t)\\
  x_{\pm}(t) &: = \bra{\psi _\pm} X(0) + Vt \ket{\psi_\pm}\\
  x_0 &:= 2 \Re [\bra{\psi_+} X(0) - (2iH)^{-1}F \ket{\psi_-}] \\
z(t) &:=  
2 \Re [\bra{\psi_+} (2iH)^{-1} e^{2iHt}F\ket{\psi_-}], 
  \end{aligned}
\end{align}
and we see that interference between positive and
negative energy states causes a shift $x_0$ of the
mean value of the position and an oscillating term
$z(t)$.
Let us now consider states whose particle and
antiparticle components are both smoothly peaked
around some momentum eingestate, i.e.
\begin{align}
  \label{eq:9}
\ket{\psi}_{in}&=c_+\ket{\psi_+}+c_-\ket{\psi_-},\\\quad \ket{\psi_\pm}&= \int
  \frac{\text{d}k}{\sqrt{2\pi}} g(k)
  \ket{\pm}_k\ket{k} 
\end{align}  
where $ |c_+|^2 + |c_-|^2 = 1 $ and $ |g(k)|^2 $ is peaked around $k_0$.
Therefore, for small value of $t$, the oscillating terms can
be approximated as follows: 
\begin{align}
  \label{eq:10}
  \begin{aligned}
    z(t) = |c_+||c_-||f(k_0)|
    \cos(2\omega(k_0)t +\phi_0)
\end{aligned}
\end{align}
where we defined 
$  f(k) \coloneqq  \bra{+_k} F \ket{-_k} /(2i\omega(k))$
and $\phi_0$ is a suitable phase.

\begin{figure*}[t!]
    \centering
    \includegraphics[width=\textwidth]{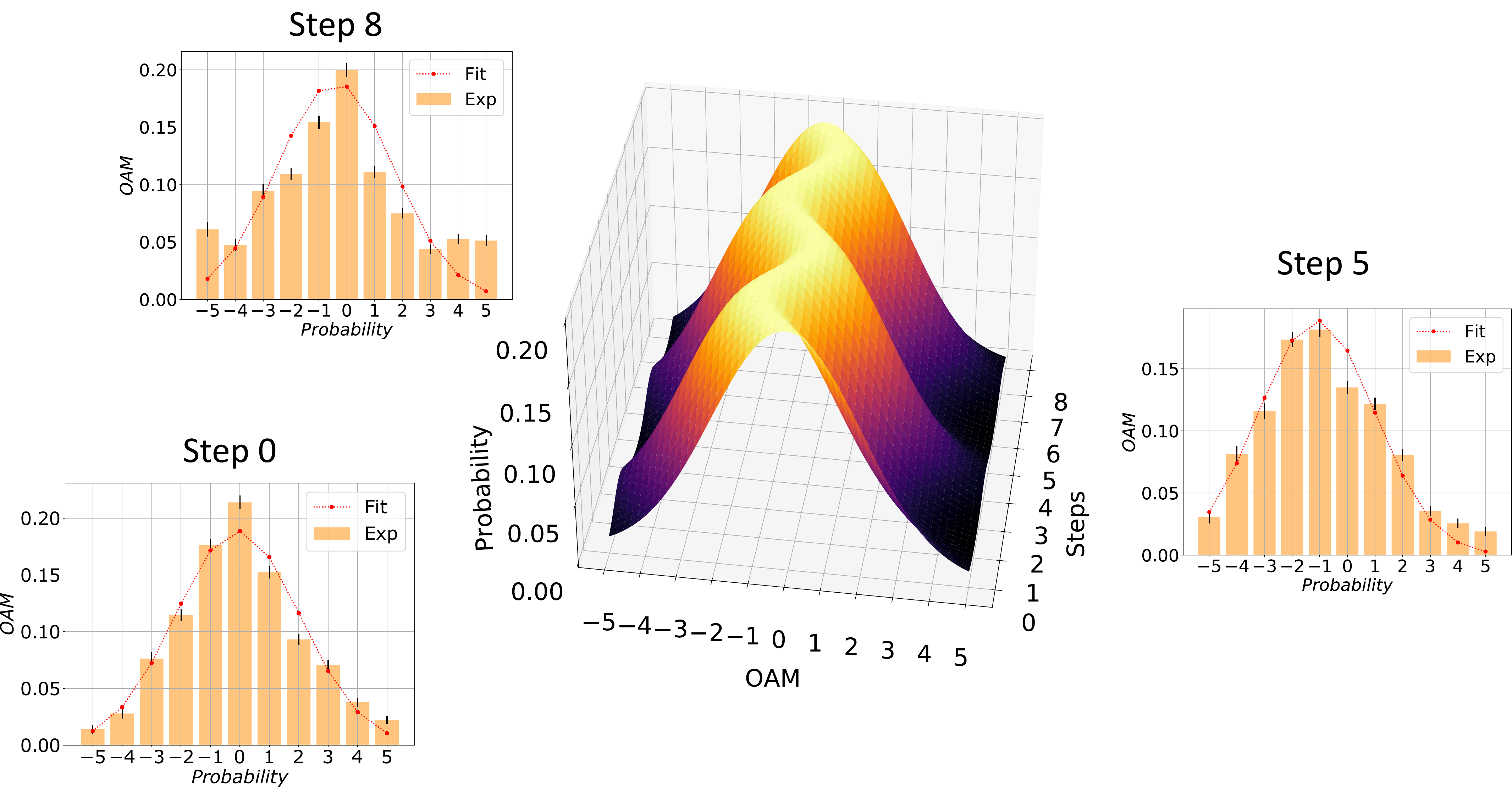}
    \caption{\textbf{Data analysis:}
    Representation of the Gaussian fit performed on experimental data. The 3d function shown is obtained by fitting the experimental data with the function in Eq. \ref{eq:fit}, where the assumed theoretical model is characterized by a Gaussian distribution that oscillates around the initial position during the evolution. In side panels, the comparison between the experimental distribution and the fitted function is reported for three different steps of the evolution (here 0 represents the input state). Although satisfactory similarities can be observed, the difference between histograms and plotted curves increases with the step evolution and this is mainly due to experimental imperfections. The reported errors on experimental data are due to the Poissonian statistics of the measured counts.
    }
    \label{fig:Fit}
\end{figure*}

\section{Experimental implementation of the Dirac Cellular Automaton}
\label{sec:3}
To experimentally realize the DQCA,
we employ the two components of photons  angular
momentum, the spin and the orbital angular
momentum, to encode coin and walker states of a quantum walk,
respectively.
The orthonormal basis $\{ \ket{R}, \ket{L}\} $
correspond to
right and left
circularly polarization, respectively.
The position states $\{ \ket{x}, x \in \mathds{Z} \}$ 
represent eingenstates of the
OAM, in particular throughout the paper we
consider its expression in the eigenstates basis
of Laguerre-Gaussian modes \cite{allen_0AM_1992}.

In our platform, the polarization
can be controlled by a set of
waveplates. 
In the circular polarized basis $\{ \ket{R}, \ket{L}\}$,
the action of a quarter-waveplate followed by a half-waveplate can be
described by the following unitary matrix:
\begin{align}
  \label{eq:12}
  C &= \frac{1}{\sqrt{2}}
      \begin{pmatrix}
        e^{2i (\alpha - \beta)} & i e^{2i \alpha} \\
        i e^{-2i \alpha} & e^{-2i  (\alpha - \beta)} 
    \end{pmatrix}  
\end{align}
where $\alpha$, $\beta$ are the angles of the
fast-axes with respect to the horizontal axis.

A conditional shift in
OAM (i.e. in the spatial degree of freedom) is
implemented using a device called q-plate
\cite{marrucci2006optical}. The latter is a
thin plate made of a birefringent material with a
direction for the optical axis that is not uniform
over the device. The angle between the optical
axis and the horizontal axis of the device follows the
relation $\gamma = \alpha_{0}+q\phi$, where
$\alpha_{0}$ is the initial angle, $q$ is the
topological  charge of the device and $\phi$ is
the azimuthal angle on the device plane. The delay
introduced on the propagation by such arrangement of the optical axis produces a modulation of the wavefront, the q-plate  action, in the momentum representation, can be described by the following
unitary operator~\cite{marrucci2006optical}:
\begin{align}
  \label{eq:11} 
  Q(k) &= \begin{pmatrix}
        \cos \frac{\delta}{2} & i e^{i 2 \alpha_0}
          \sin \frac{\delta}{2}  \: e^{ik} \\
          \\
          i e^{-i 2 \alpha_0}  \sin \frac{\delta}{2} \: 
          e^{-ik} & \cos\frac{\delta}{2}
    \end{pmatrix} ,
\end{align}
Where $k = 2q\phi$ and $\delta \in [0, \pi]$
is the q-plate tuning. 
The latter is directly
linked to the efficiency of the device in the
manipulation of the angular momentum of light,
this parameter can be electrically tuned to switch
on ($\delta = \pi$) or switch off ($\delta = 0$)
the device and, thus, control its action.\\
We realize a 8-step DTQW on a line, where each step is composed of a q-plate and a set of half-waveplate (HWP) and
quarter-waveplate (QWP). Then, the single step is given by
the composition:
\begin{align}
  \label{eq:13}
  \begin{aligned}
  U(k) = Q(k)C .
  \end{aligned} 
\end{align}
The entire setup is
enclosed between two Spatial Light Modulators
(SLMs) as shown in figure \ref{fig:QWP}-(a), a
configuration that has been already proved
suitable for the implementation of the DTQW dynamics
\cite{giordani_2018, cardano2015quantum,
  suprano2021dynamical}. The inputs of the setup
are triggered single-photon states produce via
Spontaneus Parametric Down Conversion (SPDC) in a
Periodically Poled Potassium Titanyl Phosphate
(PPKTP) nonlinear crystal. These are coupled into a
Single Mode Fiber (SMF) and then sent to the first
SLM. The latter is used to modulate the
spatial profile of photons to obtain the desired
initial state at the entrance of the quantum walk.
Therefore the input states of the setup are of the following
factorised form:
\begin{align}
  \label{eq:14}
  \ket{\psi}_{\rm{in}} = \frac{1}{\sqrt{2}}
  \Big( \ket{R} +  \ket{L}\Big) \otimes
  \sum_{x \in \mathds{Z}} g(x) \ket{x},
\end{align}
where 
$g(x) \in \mathds{R}$
  and $\sum_{x} |g(x)|^2 = 1$.
  
A second SLM instead is employed in the
measurement stage along with a single mode fiber
to project the output state onto the computational
basis and extract the occupation probability of
each OAM mode
\cite{bolduc2013holo,mair2001entanglement,forbes2016creation,Qassim:14,bouchard2018measuring,suprano2021enhanced}.
Before doing that, the polarization degree of
freedom is traced out using a series composed of a
QWP, a HWP and a Polarizing Beam Splitter (PBS).
In this way, we are able to measure only the OAM
components of the walker state at the end of the
DTQW.\\
The discrete size of SLM pixels determines the modulation efficiency of the device especially for high OAM values \cite{bolduc2013holo, Qassim:14}, while the divergence of the OAM modes 
\cite{suprano2021enhanced, cardano2015quantum} needs to be engineered and accounted depending on the number of steps. 
In particular, our setup gives us full control over OAM states $\ket{x}$
such that $|x| \leq 5$, and we choose a wavepacket which stays
confined therin for the whole evolution.
Moreover, since the platform performs up to eight evolution steps,
it is convinient that the \emph{Zitterbewegung} period $T = 2\pi/2\omega_0$, see Eq.~\eqref{eq:10},
be of the order of four, so as to observe two complete oscillations.

Let us consider the quantum walk step
\begin{align}
\label{eq:qw_dirac_evo}
	U(k) = \frac{1}{\sqrt2} \begin{pmatrix}
		e^{ik} & e^{ik} \\
		-e^{-ik} & e^{-ik}
	\end{pmatrix} ,
\end{align}
that can be experimentally implemented by choosing the following parameters $\delta = \pi$, $\alpha_0 = \pi/4$, $\alpha = -\pi/4$ and $\beta = \pi / 4$ in Eq.~\eqref{eq:13}.
One can show that the dispersion relation $\omega(k)$ of $U(k)$ is equivalent to that characterizing Eq.~\eqref{eq:2} for $m = n = \sqrt2/2$.
We are then interested in those states that are superposition of positive and negative energy eigenstates, and at the peak angular wavenumber $k_0$ feature: (i) zero group velocity $\omega'(k_0) = \partial_k \omega(k_0) = 0$, (ii) angular frequency equal to $\pi/4$, and (iii) appreciable \emph{Zitterbewegung} amplitude given by $|c_+| = |c_-| = 1/\sqrt2$ and $|f(k_0)| = 1$, see Eq.~\eqref{eq:10}.
We selected the initial state
\begin{align}
	\ket{\psi}_{\rm{in}} =
	\frac{1}{\sqrt{2}} \Big(\ket{R} + \ket{L}\Big) \otimes
	\sum_{x \in \mathds{Z}} G_{x_0, \sigma} (x) \ket{x},
\end{align}
where $G_{x_0, \sigma}(x)$ is the truncated normal distribution between $-5$ and $5$, centered in $x_0 = 0$, and with standard deviation $\sigma = 3.0$. 
For such a spatial distribution, the wavefunction in momentum representation resembles a normal distribution peaked at $k_0 = 0$ and with standard deviation $1/\sigma = 1/3$. 

This setup allows us to have a precise control and
reproduce the Dirac evolution step-by-step simply
turning on the right number of q-plates.



\section{Results}
\label{sec:4}


\begin{figure*}[ht!]
\begin{minipage}[b]{\columnwidth}
\subfloat[Ideal distribution]{\includegraphics[width=0.9\textwidth]{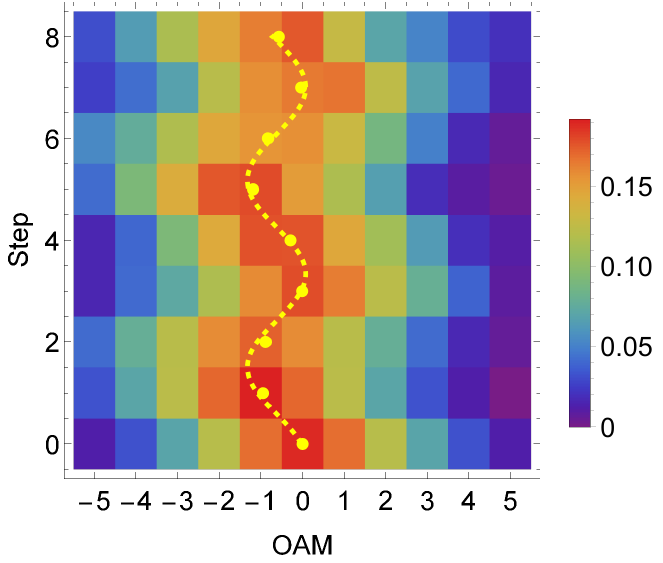}
    \label{fig:th_fig}}
\end{minipage}
\begin{minipage}[b]{\columnwidth}
\centering
\subfloat[Experimental distribution]{\includegraphics[width=0.9\textwidth]{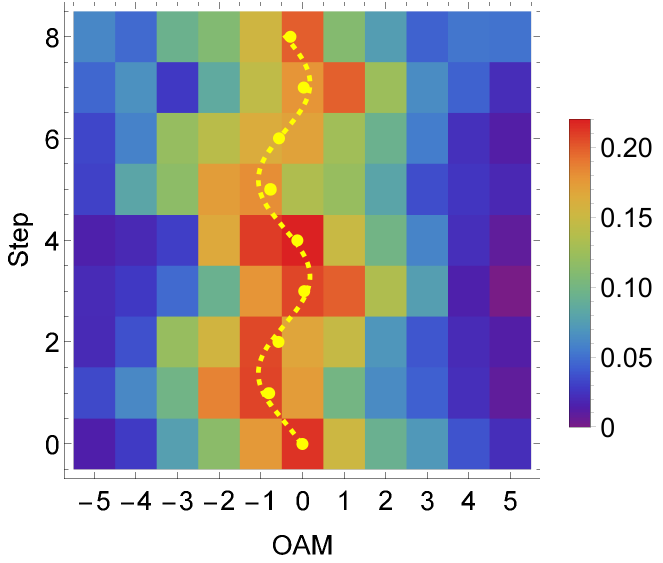}
    \label{fig:exp_fig}}
\end{minipage}
\caption{\textbf{Zitterbewegung dynamics:} The plots show the output state distribution over the OAM computational basis for each time step considered, we indicate with $0$ the initial input state. In (a) is reported the evolution obtained following the ideal noiseless model of the quantum walk, instead in (b) experimental data are shown. Yellow points represent the behavior of the mean position during the steps of the evolution while the dashed line is obtained as the step-dependent mean values of the fitted Gaussian functions.}\label{fig:ZB}
\end{figure*}

Exploiting the DTQW dynamics implemented with the setup, we 
 are able to realize a photonic QCA that allows us to experimentally study the Zitterbewegung effect of the Dirac relativistic evolution in the space of single-photon OAM. To this aim, we use q-plates with topological charge $q=1/2$ and select the angles of the waveplates in order  
to reproduce the evolution operator reported in Eq. \ref{eq:qw_dirac_evo}. Notably, we realized a state-of-the-art platform able to reach 8 steps of the DQCA evolution for arbitrary initial states in dimension 11.

We simulated the oscillatory behavior of the position of a one dimensional relativistic particle encoding this degree of freedom in the OAM of photons. Making thus the relation $\ket{x}=\ket{m}$, where $x$ is the position ad $m$ is the value of the OAM. 
This encoding is explicitly reported in Fig. \ref{fig:QWP}-b, while a conceptual representation of the simulation approach implemented in this work is shown in Fig. \ref{fig:QWP}-c.
We considered as input a Gaussian state localized around the position $\ket{0}$, generated using the first SLM in Fig. \ref{fig:QWP}-a, and observed its evolution step by step. In particular, for each step we turned on the relative q-plate setting $\delta=\pi$, traced out the information stored in the polarization and measured via the second SLM and the SMF the walker state distribution over the computational basis $\{\ket{i}\}_{i=-5}^{5}$, opportunely taking into account for the efficiencies of the measurement holograms \cite{bolduc2013holo, Qassim:14}. From the measurements, we extracted the occupation probabilities of each site and derived the evolution of the mean position (see Appendix \ref{App:noisy_model} for the measured distribution at each step and the corresponding theoretical model).   
In particular, from a theoretical prospective, we expect a Gaussian distribution that oscillates around the position $x=0$ during the evolution. The oscillation of the Gaussian peak follows the sinusoidal expression in Eq. \ref{eq:10} with frequency $\omega=2\omega(k_0)=\pi/2$ and amplitude $A=|c_+||c_-||f(k_0)|=0.5$. Since, in the experiment, we only have access to the portion of the distribution between $x=-5$ and $x=5$, the reference values for $\omega$ and $A$ are different. Therefore, at each step, we performed a fit over the distributions in a truncated interval of the position space spanned by $x \in [-5,5]$ with Gaussian functions whose mean values oscillate along the evolution direction:
\begin{eqnarray}\label{eq:fit}
    f(t,y)=\frac{e^{-(y-\mu_0 -A\cos{(\omega t+\phi)})^2/(2\sigma^2)}}{\sigma \sqrt{2 \pi}}
\end{eqnarray}
where $t$ represents the step of the DTQW, $y$ the values of probability distributions over the OAM basis, $\mu_0$ the mean of the Gaussian distribution and $\sigma$ its standard deviation. This fitting procedure is used to derive the oscillation parameters for both theoretical and experimental distributions. The results in the experimental case are shown in Fig. \ref{fig:Fit}, where the 3d plot reports the time evolution of the fitted Gaussian envelopes.

The complete experimental results are reported in Fig. \ref{fig:ZB} together with the theoretical ideal distributions. The yellow dashed lines represent the oscillations of the mean values of the fitted Gaussian functions. For the theoretical distribution the sinusoidal curve in Fig. \ref{fig:th_fig} is characterized by values equal to  $\omega= 1.714 \pm 0.017$ and $A=0.695 \pm 0.032$.
Experimentally, we obtained an oscillation very similar to the expected one with values 
that correspond to $\omega= 1.655 \pm 0.009$ and $A=0.615 \pm 0.017$, the measured behavior is reported in (Fig. \ref{fig:exp_fig}).
From both numerical results and plots shown in Fig. \ref{fig:ZB} it can be seen how the implemented platform is capable of simulating the dynamics of a free relativistic particle, reproducing its typical Zitterbewegung trembling motion.  



\begin{figure*}[ht!]
    \centering
    \includegraphics[width=\textwidth]{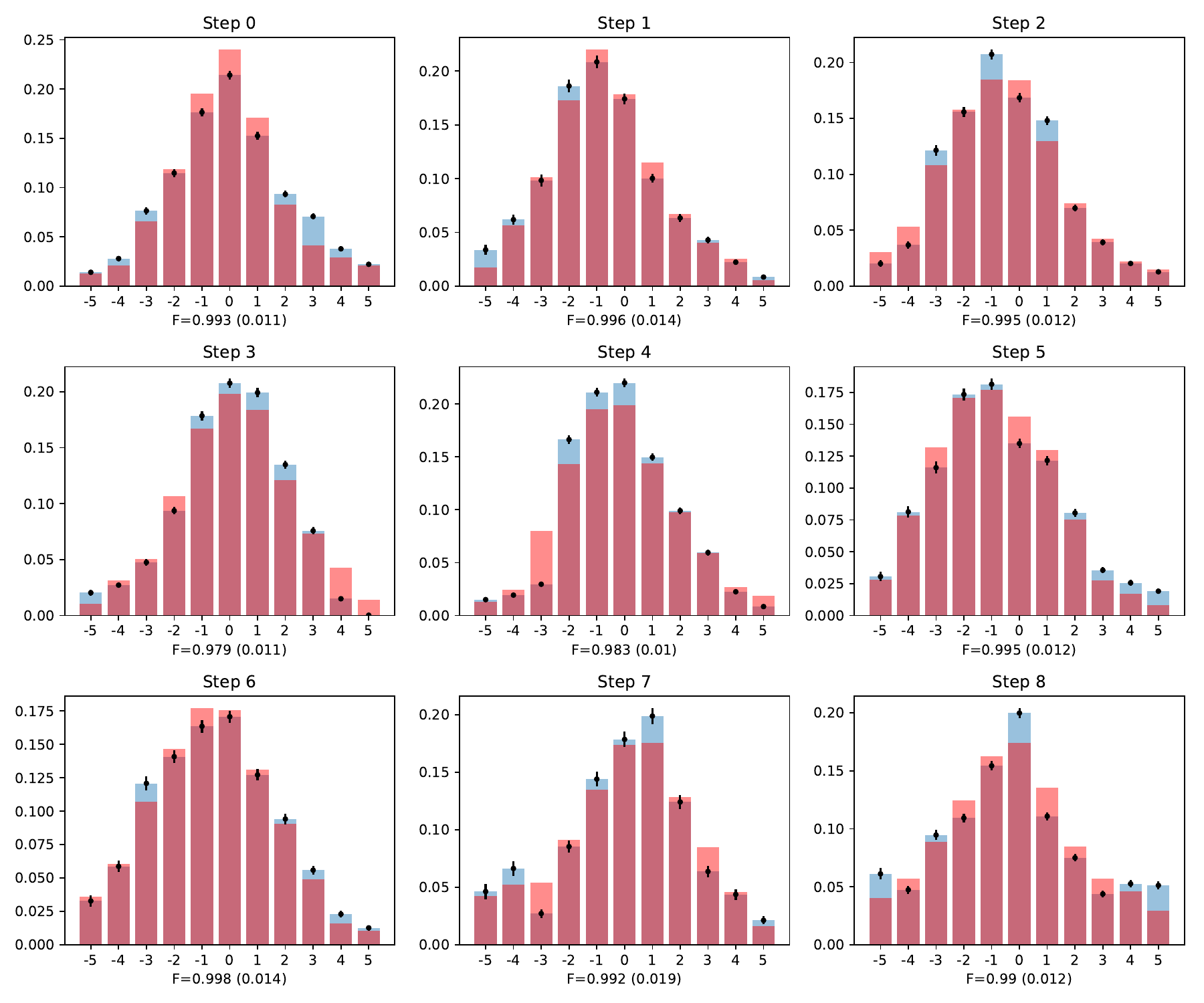}
    \caption{\textbf{Step by step QCA evolution:} OAM occupation probability measured at the output of the setup for different steps of the DTQW dynamics, above each histogram the relative step is indicated. The experimental data are reported in blue, while in red are reported the expected behaviors from the theoretical noisy model of the experimental apparatus. The fidelity between the experimental and theoretical distributions is reported under each histogram. Error bars are due to the Poissonian statistics of single photons counting.}
    \label{fig:isto}
\end{figure*}

\section{Conclusions}


We showcase the photonic implementation of a Quantum Cellular Automaton able to simulate features typical of the Dirac free particle evolution. Our work makes a relevant step forward the exploration of experimental QCA, and especially DQCA, in photonic platforms.
The single-particle sector of the DQCA is realized through the discrete time quantum walk dynamics performed exploiting the OAM of single photons. In particular, the Quantum Walk platform is composed of a cascade of 8 q-plates interspaced by waveplates and placed between 2 SLMs. This setup,   advancing the state-of-the-art OAM based DTQW platforms, allowed us to have an high control over the input state and the capability to perform 8 steps of the automaton evolution. The simulation power of the DQCA was used to reproduce the Zittebewegung, i.e. the trembling motion that occurs during the free evolution of relativistic particles. These effects are clearly reproduced in the reported results, in which the particle position, encoded in the OAM value of single photons, presents an oscillatory behavior with an amplitude and a frequency in agreement with the theoretical predictions.
Moreover, these experimental results represent a proof of principle demonstrating the possibility of employing photonic platforms to implement DQCAs. The proposed experimental protocol constitutes a first step towards the simulation of more complex dynamics where position-dependent evolutions are necessary, such as the Dirac particle evolution subject to external potential \cite{Cedzich2013Propagation,cedzich2019quantum} and curved spacetime \cite{Molfetta2013Quatum,arrighi2016quantum,mallick2019simulating}. On the other side,  the potential of the implemented DQCA can be extended to the multiparticle regimes for instance, by considering entangled photons in parallel quantum walks \cite{giordani2021entanglement}. Although the DTQW dynamics is a special case of QCA evolution, the provided photonics tools are completely general and thus can aim at the implementation of more general multidimensional QCA in suitable experimental photonics platforms.

\section*{Acknowledgments}
PP and FS acknowledge financial support from European Union – Next Generation EU through the MUR project Progetti di Ricerca d'Interesse Nazionale (PRIN) QCAPP No. 2022LCEA9Y. FS also acknowledge support from the Templeton Foundation, The Quantum Information Structure of Spacetime (QISS2) Project (qiss.fr) (the opinions expressed in this publication are those of the author(s) and do not necessarily reflect the views of the John Templeton Foundation)  Grant Agreement No. 62312. AB acknowledges financial support from European Union – Next Generation EU through the PNNR MUR project PE0000023.
\appendix
\section{Noisy theoretical model and experimental data}
\label{App:noisy_model}
In this section, we present in more details the experimental result of the QCA evolution together with the theoretical model used to account for the experimental imperfections present in the setup. These are mainly related to the non unitary efficiencies of the cascaded q-plates, errors in the coin operators and the efficiencies of the holographic technique in manipulating the OAM \cite{bolduc2013holo, Qassim:14}. 
In this way, the DTQW evolution is simulated by means of a theoretical model whose parameters are the conversion efficiencies of the q-plates, 
the offset of the waveplates 
and the efficiencies of the two SLMs. 
These parameters are optimized around the attended values by simultaneously  minimizing the square difference between the theoretical evolution and the raw measured data for each step. 
All the experimental distributions, measured at each step, are shown in Fig. \ref{fig:isto}. Here, it is reported the OAM probability occupation as expected from the noisy theoretical model and the experimental unfolded one. The similarity between the theoretical and experimental distributions is quantified by the fidelity value reported under each histogram.

\bibliographystyle{apsrev4-2}

\end{document}